\documentclass[12pt,preprint]{aastex}

\usepackage{amsfonts}
\usepackage{amsmath}
\usepackage{amssymb}
\usepackage{wrapfig}
\usepackage{graphicx}
\usepackage{bibentry,natbib}
\usepackage{wasysym}
\usepackage{latexsym}
\usepackage{subfigure}
\usepackage{breqn}

\newfont{\myfont}{cmmib10}

\shorttitle{Changes in Polarization Angle Across the Eclipse in the
  Double Pulsar System} 
\shortauthors{authors}

\begin{document}

\title{Changes in Polarization Position Angle across the Eclipse in the Double Pulsar System}

\author{R. Yuen\altaffilmark{1,2}, R. N. Manchester\altaffilmark{1}, 
M. Burgay\altaffilmark{3}, F. Camilo\altaffilmark{4}, M. Kramer\altaffilmark{5}, 
D. B. Melrose\altaffilmark{2} and I. H. Stairs\altaffilmark{6}}

\affil{\altaffilmark{1}CSIRO Astronomy and Space Science, Australia
  Telescope National Facility, P.O. Box 76, Epping, NSW 1710,
  Australia} 
\affil{\altaffilmark{2}SIfA, School of Physics, the
  University of Sydney, NSW 2006, Australia}
\affil{\altaffilmark{3}INAF-Osservatorio Astronomica di Cagliari, Loc. Poggio dei Pini, Strada 54, 09012 Capoterra, Italy} 
\affil{\altaffilmark{4}Columbia University, New York, NY 10027, USA}
\affil{\altaffilmark{5}Max Planck Institut f$\ddot{u}$r Radioastronomie, Auf dem H$\ddot{u}$gel 69, 53121 Bonn, Germany} 
\affil{\altaffilmark{6}Department of Physics and Astronomy, University of 
British Columbia, Vancouver, BC V6T 1Z1, Canada}

\begin{abstract}

  We investigate the changes in polarization position angle in
  radiation from pulsar A around the eclipse in the Double Pulsar
  system PSR J0737-3039A/B at the $20\,{\rm cm}$ and $50\,{\rm cm}$ wavelengths using the
  Parkes 64-m telescope. The changes are $\sim 2\sigma$ during and
  shortly after the eclipse at $20\,{\rm cm}$ but less significant at $50\,{\rm cm}$. We
  show that the changes in position angle during the eclipse can be
  modelled by differential synchrotron absorption in the eclipse
  regions. Position angle changes after the eclipse are interpreted as
  Faraday rotation in the magnetotail of pulsar B. Implied charge
  densities are consistent with the Goldreich-Julian density,
  suggesting that the particle energies in the magnetotail are mildly
  relativistic.
\end{abstract}

\keywords{binaries: eclipsing --- polarization --- pulsars: individual
  (PSR J0737-3039A, PSR J0737-3039B)}

\section{Introduction}

PSR J0737-3039A/B, also known as the Double Pulsar system, is a
compact binary system consisting of two neutron stars, both radiating
at radio frequencies\footnote[1]{PSR J0737-3039B is currently
  undetectable at radio frequencies, probably because of relativistic
  precession of its spin axis \citep{Perera2010}.}. Discovered by
\citet{Burgay2003} and \citet{Lyne:etal2004} in a survey at $20\,{\rm cm}$
wavelength using a multibeam receiver on the 64-m radio telescope at
Parkes in New South Wales, Australia, the two pulsars, referred to as
pulsar A and pulsar B respectively, orbit each other with a period of
about 2.45 hours in an almost circular trajectory with eccentricity
0.088 \citep{Burgay2003, Kramer:etal2006}. The inclination angle
between the line of sight and orbital plane normal is estimated to be
$87.7^{\circ}$. The spin periods for pulsar A and B are $P_A =$ 22.7
ms and $P_B =$ 2.77 s, and their surface dipole magnetic field
strengths are $B_A = 6.3\times 10^{9}\,\rm G$ and $B_B = 1.2\times
10^{12}\,\rm G$. The spin-down powers for pulsar A and B are
$\dot{E_A}\sim 5.9\times 10^{33}\,\rm{erg\,s^{-1}}$ and $\dot{E_B}\sim
1.6\times 10^{30} \,\rm{erg\,s^{-1}}$, giving $\dot{E_A}/\dot{E_B}
\sim$ 3500.

\begin{figure}
\begin{center}
\includegraphics{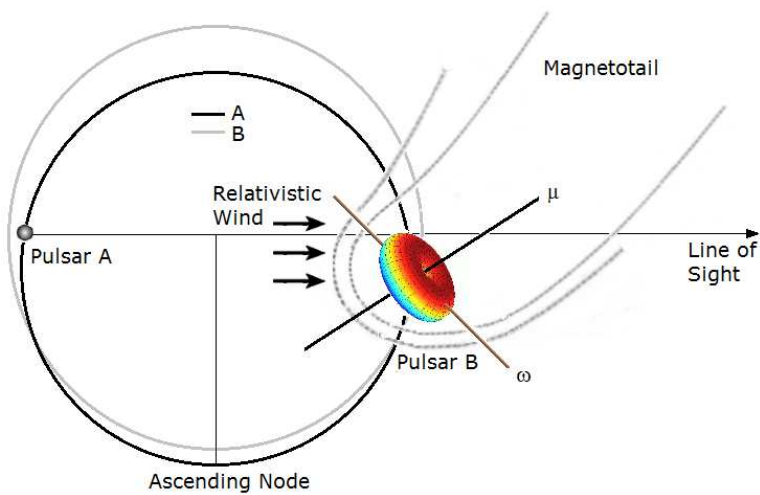}
\caption{Orbital arrangements of the two pulsars and the magnetotail
  behind pulsar B (not to scale). The black and grey curves represent
  the orbits for pulsar A and B respectively and both pulsars orbit in
  the clockwise direction. The shaded areas around pulsar B represent
  the closed field-line regions (adapted from \citet{Breton2008} and
  online material), and $\omega$ and $\mu$ are the angular velocity and
  magnetic moment vectors respectively. The orbital normal is directed
  out of the page. As the two pulsars orbit, the radiation we observe
  from pulsar A first encounters the leading edge of the shock front,
  enters the magnetosphere of pulsar B, leaves the magnetosphere on
  the other side and finally crosses the magnetotail.}
\label{fig-OrbitalMagnetotail}
\end{center}
\end{figure}

Because of the strong relativistic wind from pulsar A, the
magnetosphere of pulsar B is truncated compared to that of an isolated
pulsar. The relativistic particle flux in the wind penetrates into the
light cylinder of B resulting in the formation of a bow shock that
strongly perturbs the dipolar field structure by compressing B's
magnetosphere on the ``daytime'' side (the side facing A), and
stretching it on the ``nightime'' side into a magnetotail. From
studies on simulating the interaction between A's wind and B's
magnetosphere \citep{Arons2004}, the magnetotail probably reaches a
distance of several semi-major axes of pulsar B's orbit (the
semi-major axis is $4.5\times 10^8$ m). Figure
\ref{fig-OrbitalMagnetotail} illustrates the orbital
arrangements for the two pulsars and the magnetotail. The wind from
pulsar A blows at pulsar B's magnetosphere causing B's magnetospheric
material to flow in the radial direction, but because of the orbital
motion, the magnetotail is curved behind pulsar B.

The fact that the orbit is almost edge-on to the line of sight makes
an eclipse observable each orbit when pulsar A moves behind pulsar
B, as shown in Figure \ref{fig-OrbitalMagnetotail}. The eclipse occurs
at orbital longitude of $\sim 90^\circ$ measured from the ascending
node and lasts for $\sim$ 30 seconds \citep{Lyne:etal2004}. It was shown that, during an eclipse, the
pulsed flux density of A is modulated at the rotational period of pulsar B and its second harmonic
\citep{McLaughlin:etal2004}. Synchrotron absorption by the highly
relativistic plasma in the closed field-line regions of pulsar B has
been proposed as the eclipse mechanism \citep{LyutikovThompson2005}. Modelling of these variations by \citet{Breton2008} gave best-fit values of $\sim 130\degr$ for the angle between rotation axis and the orbital normal, and $\sim
70\degr$ for the inclination angle between the rotation and magnetic axes. Hence pulsar A's radiation passes through pulsar B's
truncated magnetosphere and interacts with the plasma during an
eclipse, then, as the two pulsars advance in orbit away from the
superior conjunction, pulsar A's radiation begins to traverse the
magnetotail.

The radiation from pulsar A that penetrates the magnetosphere of
pulsar B during and shortly after an eclipse serves as a probe into
the plasma content of B's magnetosphere and magnetotail. Study of the
emergent radiation may reveal the largely unknown properties of pulsar
magnetospheres and possible physical processes within them. In this paper,
we investigate the changes in polarization position angle in radiation
from pulsar A at $20\,{\rm cm}$ and $50\,{\rm cm}$ wavelengths during and around the
eclipses.

\section{Observation and data analysis}

Our observations of the Double Pulsar use the Parkes 64-m radio
telescope. The data of interest were taken in two frequency bands
centred at 1369 MHz ($20\,{\rm cm}$) from September 2007 to October 2011, and
at 730 MHz ($50\,{\rm cm}$) from February 2010 to November 2011. The $20\,{\rm cm}$
observations were made using the center beam of the $20\,{\rm cm}$ Multibeam
receiver \citep{swb+96} and the $50\,{\rm cm}$ observations used the $10\,{\rm cm}$/$50\,{\rm cm}$
dual-band receiver \citep{gzf+05}. The Parkes digital filterbank
systems, PDFB3 and PDFB4, were used to split the $20\,{\rm cm}$ and $50\,{\rm cm}$ data
into 1024 and 512 channels respectively, form the polarization products
and fold the data at the topocentric pulse period. All observations
had a sub-integration time of $30\,{\rm s}$. We checked the presence of an
eclipse in a data file by checking the coverage of orbital phase of
$90^\circ$. Once we had identified all relevant data files, regions $10\,{\rm min}$ before and after the eclipse were retained for further
analysis using the {\sc psrchive} software suite \citep{HSM2004}. Spectral
band edges were deleted and interference was removed from all
files. Files that had significant interference at or close to the
eclipse region were eliminated from our list. Short observations of a
linearly polarized broad-band calibration signal preceding each pulsar
observation were used to calibrate the instrumental phase and
gain. For the $20\,{\rm cm}$ system, the effects of cross-coupling between the
orthogonal feed probes were removed as part of the calibration
process. Data were summed in frequency using the corrected rotation
measure, $+112.3\, {\rm rad\, m}^{-2}$ \citep{MKS+10}. We obtained 38
files at $20\,{\rm cm}$ and 13 at $50\,{\rm cm}$ with good quality data.

To improve the signal-to-noise ratio, we summed the files in each
band, binning the results in orbital phase. The phase bins were $\sim
0.7\degr$ or $\sim$ $18\,{\rm s}$ in width. Sub-integration data were assigned to
the two adjacent orbital phase bins with weights proportional to
$1-x$ and $x$ for the leading and trailing bins respectively,
 where $x$ is the interval between the leading phase bin centre and the sub-integration center time, normalized by the phase bin
width. Finally, Stokes parameters were summed across the main pulse and
interpulse to form the average Stokes parameters for each orbital
phase bin. This summing of data across the two pulses is possible
because the strongly linearly-polarized parts of both pulse components
have essentially constant and equal position angle \citep{drb+04}.

\section{The results}
\label{sec:TheResults}

\begin{figure}
\begin{center}
\includegraphics{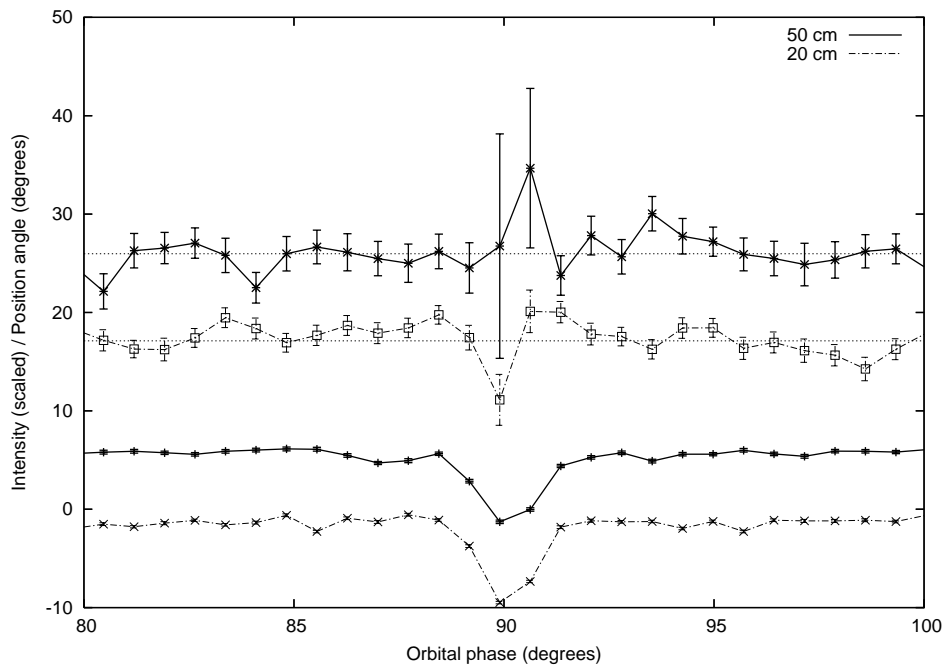}
\caption{Plot of polarization position angle in degrees (upper curves)
  and Stokes I (total intensity) parameter (lower curves) plotted
  against orbital phase in degrees for $50\,{\rm cm}$ (solid curve) and $20\,{\rm cm}$
  (dashed curve) at $\sim18$~s time resolution. The curve for position
  angle at $50\,{\rm cm}$ is offset by $3^\circ$ for clarity. The Stokes I
  curves clearly show the
  eclipse centred near $90^\circ$. The horizontal lines represent average baseline values.}
\label{fig-PA_Plot}
\end{center}
\end{figure}

Measurements of the mean position angles and Stokes I (total
intensity) for $20\,{\rm cm}$ and $50\,{\rm cm}$ wavelengths as a function of orbital
phase are presented in Figure \ref{fig-PA_Plot}. Uncertainties in the
position angle for each orbital phase were computed using $\sigma =
28.65^\circ (\sigma_I / L_c)$ \citep{HandbookOfPulsarAstronomy}, where
$\sigma_I$ is the standard deviation of the total intensity, and $L_c$
is the linear polarization corrected for noise bias. The Stokes I
curves, which are scaled to fit into the plot, clearly show a large
drop in intensity representing the eclipse region.

\begin{figure}
\begin{center}
\includegraphics{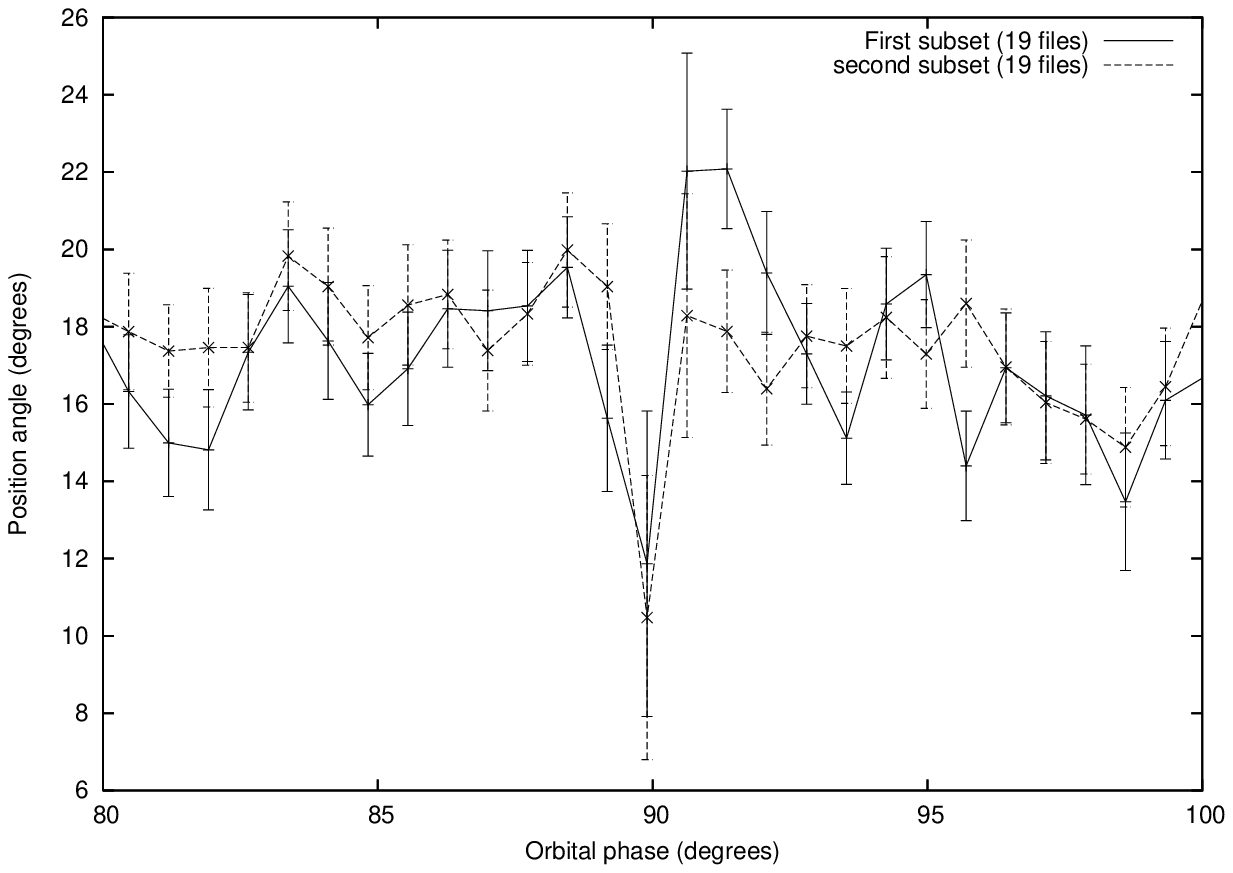}
\caption{Position angles at $20\,{\rm cm}$ for two subsets of the total data
  set plotted against orbital phase.}
\label{fig-TwoSubsets20cm}
\end{center}
\end{figure}

We consistently detected changes in position angle during and shortly
after an eclipse at both wavelengths. Inspection of the $20\,{\rm cm}$ curve in
the eclipse region shows that the position angle decreases during the
ingress phase with a change of approximately $6^\circ \pm 3^\circ$
relative to the mean baseline level. It then rises to its maximum
value of approximately $3^\circ \pm 2^\circ$ above the baseline at the
end of the eclipse. It remains high for the following phase bin, then
drops back to the baseline value. There are also changes in
polarization angle in the corresponding regions during the eclipse at
$50\,{\rm cm}$, but the uncertainties are larger. Nevertheless, the position
angle changes shortly after the eclipse is consistent with those at $20\,{\rm cm}$. 

We confirmed our results by randomly partitioning the $20\,{\rm cm}$ files into two
subsets as shown in Figure \ref{fig-TwoSubsets20cm}. The changes in position angle
during the eclipse are essentially consistent between the two subsets,
i.e., a decrease during ingress followed by an increase during
egress. Partitioning the dataset into either two, three or four subsets,
either randomly or chronologically, produced similar results.

\section{Mechanisms to change the polarization position angle}

\subsection{\it Synchrotron absorption}

Synchrotron absorption, like synchrotron emission, may be separated
into two linearly polarized components, with the more strongly emitted
and absorbed component having its electric vector perpendicular to the
magnetic field. The polarization-averaged synchrotron absorption
coefficient for a relativistic thermal distribution of particles,
$n(\varepsilon) \propto e^{-\varepsilon/k_BT}$, is
\citep{LyutikovThompson2005},
\begin{equation} \label{eq:SynchrotronAbsorptionCoeff} \alpha_\nu =
  \frac{4\pi^2en_s}{3^{7/3}B\, {\rm sin}\, \kappa}\, \Bigg[ \,\Bigg(\,
  \frac{m_ec^2}{k_BT} \,\Bigg)\, \frac{\nu_{{\rm B},e}{\rm sin}\,
    \kappa}{\nu}\, \Bigg]^{5/3},
\end{equation}
for $\nu \ll \nu_p \sim 4(k_B T_e/m_e c^2)^2 \nu_{B,e}$, where
$n_s$, $B$, $m_e$, $c$, $k_B$, $T$, and $\nu$ are the density of the
absorbing particles, magnetic field strength, electron mass, speed of
light, Boltzmann constant, plasma temperature and radiation frequency
respectively, $\kappa$ is the angle between the magnetic field and
propagation vector of the radiation and $\nu_{B,e} = eB/2\pi m_ec$ is
the cyclotron frequency. The absorption coefficients for polarization
perpendicular and parallel to the magnetic field are $\alpha_\nu^\perp
= 4\alpha_\nu/3$ and $\alpha_\nu^\parallel = 2\alpha_\nu/3$, and so
$\alpha_\nu^\perp/\alpha_\nu^\parallel = 2$. The optical depth is
$\tau_\nu = \alpha_\nu L$, where $L$ is the effective path length
through the absorbing medium. Hence, synchrotron absorption
preferentially removes the linear polarization perpendicular to the
magnetic field. As a result, the position angle of the transmitted
radiation changes with increasing optical depth, with the position
angle approaching the angle of the projected magnetic field. If
$\chi_0$ and $\chi_\nu$ are the incident and emergent polarization
angles relative to the projected magnetic field direction, then,
\begin{eqnarray} \label{eq:IntensityChange} I_\nu^\perp = I_\nu\, {\rm
    sin}\, \chi_\nu = I_{0,\nu}^\perp\, {\rm exp}(-\tau_\nu^\perp) =
  I_{0,\nu}\, {\rm sin}\, \chi_0\, {\rm exp}(-4\tau_\nu/3)
  & \nonumber \\
  I_\nu^\parallel = I_\nu\, {\rm cos}\, \chi_\nu =
  I_{0,\nu}^\parallel\, {\rm exp}(-\tau_\nu^\parallel) = I_{0,\nu}\,
  {\rm cos}\, \chi_0\, {\rm exp}(-2\tau_\nu/3).
\end{eqnarray}
where $I^{\perp,\parallel}$ represents intensity perpendicular and
parallel to the projected magnetic field respectively, and quantities
with subscript zero represent incident values at a particular
frequency, $\nu$. The change in position angle is given by
$\Delta\psi_\nu = (\chi_\nu - \chi_0)$.

\subsection{\it Faraday rotation}

The decomposition of linearly polarized radiation into two circular modes that propagate at slightly different phase velocities results in Faraday rotation of the plane of the received linear polarization. Integrating over the full path gives the change in position angle, $\Delta\psi$
\begin{equation} \label{eq:FaradayRotation}
\Delta\psi \approx 0.81\, \lambda^2\, \int^L_0 n_e B\, {\rm cos}\, \theta\, ds = 0.81L\, \langle n_eB \rangle\, \lambda^2 = {\rm RM}\, \lambda^2,
\end{equation}
where $n_e$, $B$ and $L$ are the electron density in ${\rm cm}^{-3}$,
the average magnetic field in microgauss, and the pathlength in
parsecs respectively, and ${\rm cos}\, \theta$ is the angle between
the magnetic field and the ray path, ${\rm RM}$ is the rotation
measure. The notation $\langle...\rangle$ implies an average along the
ray path.

Pulsar magnetospheres are thought to be populated with highly
relativistic plasma. If there is no thermal plasma present, as some
models imply, the modes would be nearly linearly polarized, and
conventional Faraday rotation would not occur. The natural wave modes
would also be linear if the density of the electrons and positrons
were the same. The circular component needed can arise from the net
charge, associated with the Goldreich-Julian density, provided the
particle energies are not highly relativistic. Any measured Faraday
rotation would have the interesting implication that the dominant
contribution to the wave dispersion is from electrons that are no more
than mildly relativistic.

\section{Discussion} \label{sec:Discussion}

\subsection{\it Changes in position angle during the eclipse} \label{sec:ChangesInPolarizationAngleDuringEclipse}

Our measurements of the position angle change of A's pulse emission
integrate over the rotation of B's magnetosphere and the resulting
eclipse modulation, with our $18\,{\rm s}$ averaging time covering approximately
6.5 revolutions of B. The direction of the differential synchrotron
absorption relative to the incoming radiation will vary during B's
rotation and along the path through the magnetosphere. It will also
change as the line of sight to A moves through B's magnetosphere
because of the orbital motion. Our present results do not justify a
full computation of this complex variation. However, we can
approximate the effect by recognising that the closed field lines
perpendicular to the magnetic axis are symmetrical but of opposite
sign in the two hemispheres. As the star rotates, they will therefore
average to zero leaving a net magnetic field along the magnetic
axis. Furthermore, since the magnetic and rotation axes are not
aligned, the field perpendicular to the rotation axis will also cancel
leaving a net magnetic field along the rotation axis of B. During the
four years covered by our observations, B's spin axis will precess by
about $20\degr$ but this will have little effect of the observed
absorption which is dominated by averaging over B's rotation

The peaks and dips in flux density from pulsar A in the eclipse light
curve \citep{McLaughlin:etal2004} imply that the optical depth is,
respectively, either $\sim 0$ when the line of sight misses B's
magnetopsheric torus, or $>1$, which corresponds to the absorbing
closed-field region that obstructs our view of pulsar A resulting in a
large decline in flux density. The differential changes in
$I_\nu^\perp$ and $I_\nu^\parallel$ would be insignificant during most
of the eclipse region where $\tau > 1$ because the intensities are
low, except near the boundaries of the eclipse where the optical depth
$\tau\sim 1$, which characterizes the applicability of Equation
(\ref{eq:IntensityChange}). For the synchrotron absorption model
(Equation \ref{eq:SynchrotronAbsorptionCoeff}), the density of the
absorbing plasma can be estimated by $n_s = M\, n_{GJ}$, where $M =
10^6$ is the pair multiplicity parameter, and $n_{GJ}$ is the
\citet{Goldreich1969} charge density. The energy balance between
A's relativistic wind and B's magnetic pressure gives a field strength of 7~G at the eclipse bounary, which is estimated to be near $r_{\rm
  bal} \approx r_L/10 \sim 1.5 \times 10^9$ cm from pulsar B \citep{LyutikovThompson2005}, and so $n_{GJ} \sim 0.2\, {\rm
  cm}^{-3}$, giving $n_s \sim 2\times 10^5\, {\rm cm}^{-3}$. The
cyclotron frequency, $\nu_{B,e}$ is $2\times 10^7$ Hz, and the
relativistic electrons are assumed to be thermally distributed at
temperature $(k_BT/m_ec^2) = 10$. Assuming B's magnetic field is
randomly oriented and averaging between $0^\circ$ and $90^\circ$ gives
$\langle ({\rm sin}\, \kappa)^{2/3} \rangle \approx 2/\pi$, and with
$L = r_{\rm bal}$, we obtain $\langle\tau_{20}\rangle \sim 1$ and
$\langle\tau_{50}\rangle \sim 3$, where the number in subscript
represents the wavelength. Using these values and our data for the
total intensity, and assuming $\chi_0 = 45^\circ$, Equation
(\ref{eq:IntensityChange}) gives an order of magnitude change in
polarization or position angle of $10\degr$, consistent with our
measurements. We therefore, conclude that differential synchrotron
absorption in the closed field-line regions can plausibly account for
the observed position angle changes during the eclipse.

Combining Equation (\ref{eq:IntensityChange}) and rearranging gives
${\rm tan}\, \chi_\nu = {\rm tan}\, (\chi_0 + \Delta\psi_\nu) = {\rm
  tan}\, \chi_0\, {\rm exp}(-2\tau_\nu/3)$. In principle, with higher
time-resolution observations of the eclipse, one could use this
equation to determine the angle of the rotation axis of pulsar B
projected on the sky, $\psi_{B,\Omega}$. At times when $\tau_\nu$ is
neither very large nor very small, both $\tau_\nu$ and
$\Delta\psi_\nu$ may be measurable. Since $\chi_0 = \psi_0 -
\psi_{B,\Omega}$ and $\Delta\psi_\nu = \chi_\nu - \chi_0 = \psi_\nu -
\psi_0$, the above equation can be rewritten as ${\rm tan}(\psi_\nu -
\psi_{B,\Omega}) = {\rm tan}(\psi_0 - \psi_{B,\Omega})\, {\rm
  exp}(-2\tau_\nu/3)$.  With multiple measurements for each of
$\psi_\nu$ and $\tau_\nu$, and since $\psi_0$ is known,
$\psi_{B,\Omega}$ may be determined.

\subsection{\it Changes in position angle due to Faraday rotation}

After the eclipse, the synchrotron absorption effect will diminish and
Faraday rotation should start to manifest. During this time, pulsar
A's radiation goes through regions located immediately beyond the
closed field-line boundaries in the wind zone where the oppositely directed magnetic
fields in the two lobes from the magnetotail each connect to one of
the polar caps. Unlike the geomagnetotail where field structure is
symmetric in these regions \citep{PilippMorfill1978}, the large
inclination angle results in an asymmetric field structure between the
two lobes in these regions and an alternating field structure as the
pulsar rotates within the confining magnetosheath
\citep{Spitkovsky}. With the misalignment between the rotation axis
and the orbital normal, a net magnetic field in the line of sight will
result in Faraday rotation in radiation that traverses these regions
of pulsar B's magnetotail.

Faraday rotation causes changes in
position angle of the same sign at both wavelengths shortly after an
eclipse. The measured changes in position angle can be used to
estimate $\langle n_eB \rangle L$, and a separate estimate of $L$ and
$B$ allows $n_e$ to be estimated in the magnetotail. Consider mean
values along the line of sight, and assuming magnetic field scales as
$\propto r^{-2}$ up to $L/2$ in the magnetotail\footnote[2]{Scaling of
  the magnetic field strength in the open field-line region and
  magnetotail is uncertain but not critical to our discussion.} gives
$B \sim$ 2~G, where $L \approx 1.5 \times 10^{-8}$ pc is the
semi-major axis of pulsar B's orbit. Equation
(\ref{eq:FaradayRotation}), with $\Delta\psi_{50} \sim -5\degr$, gives
$n_e \sim 40\, {\rm cm}^{-3}$.  The corotating charge density \citep{Goldreich1969} is the
difference between the number densities of positrons and electrons,
which gives $\sim 20\, {\rm
  cm}^{-3}$ at $r = r_{\rm bal}$. Although the corotation requirement
does not apply beyond $r_{\rm bal}$, the wind in the magnetotail
rotates with the star implying a charge density of order the
Goldreich-Julian value. This suggests that Faraday rotation may be
responsible for the changes in position angle away from the eclipse
region. As mentioned earlier, the complication in interpreting the
results lies in the complex variations in the orientation of B's
magnetic field structure, caused by field reversals between
the two lobes in the magnetotail, the large inclination angle
between magnetic and rotation axes of pulsar B and the orbital motion
of the two pulsars. However, with higher time resolution, it is possible that
changes in A's position angle could be correlated with different orientations of
B's field structure.

\subsection{\it Mildly relativistic particle energies}

Faraday rotation in the magnetotail would require that the charged particles
be no more than mildly relativistic. If the particles were highly relativistic, the natural wave modes would be linearly polarized and there would be no Faraday rotation. If our suggestion that
significant Faraday rotation occurs is confirmed by more detailed
observations, it would be a strong indication that the bulk of the
pairs produced in a pulsar magnetopshere are no more than mildly
relativistic. This is inconsistent with theories for the pair creation
that suggest much higher mean Lorentz factors of typically $\gamma\sim
10^2 - 10^4$ \citep{Sturrock1971, RudermanSutherland1975,
  DaughertyHarding1982}. Since energy-loss mechanisms are ineffective
\citep{RudermanSutherland1975, Sturner1995, LyubarskiiPetrova2000},
any nonrelativistic plasma must be created directly in the pulsar
magnetosphere \citep{JessnerLeschKunzl2001}. Direct generation of
mildly relativistic pairs is favored for magnetic fields with $0.02\,
B_{\rm cr} < B < 0.1\, B_{\rm cr}$, where $B_{\rm cr} \approx 4.9
\times 10^{13}$ G is the critical field strength
\citep{WeiseMelrose2002}.

\section{Conclusions}

We have confirmed the changes in polarization position angle in
radiation from pulsar A during and shortly after the eclipse in the
Double Pulsar system at $20\,{\rm cm}$ and $50\,{\rm cm}$ wavelengths. The changes are
$\sim 2\sigma$ at $20\, {\rm cm}$ and $\sim 1\sigma$ at $50\, {\rm cm}$. The variations at $20\,{\rm cm}$ were confirmed by partitioning the
data files into two subsets. Differential synchrotron absorption in the
closed field-line regions can account for the observed position angle
changes during the eclipse. Modelling of the changes in position angle
shortly after the eclipse with Faraday rotation gives a charge density
that is consistent with the Goldreich-Julian value. Since energy-loss
mechanisms in pulsar magnetospheres are inefficient, the presence of
Faraday rotation implies that pairs must be created directly in
nonrelativistic regime.

\section*{Acknowledgments}
We thank our colleagues for assistance with the observations reported
in this paper. The Parkes radio telescope is part of the Australia
Telescope National Facility which is funded by the Commonwealth of
Australia for operation as a National Facility managed by
CSIRO. Pulsar research at UBC is supported by an NSERC Discovery
Grant.


\end{document}